# EXACT ANTICIPATING SYNCHRONIZATION IN BI-DIRECTIONALLY COUPLED TIME-DELAYED SYSTEMS

E. M. Shahverdiev [1], S.Sivaprakasam and K. A. Shore
School of Informatics, University of Wales, Bangor, Dean Street, Bangor, LL57 1UT, Wales, UK

We present an analytical investigation of exact anticipating synchronization between two bi-idirectionally coupled time-delayed systems in the case when these systems are governed by two characteristic times: the delay time in the coupling and the delay time in the coupled systems themselves. We find that exact anticipating chaos synchronization with coupling-delay anticipating times can exist if there are parameter mismatches. We apply our approach to interpret the results of recent first experimental observation of anticipating chaos synchronization in external cavity semiconductor lasers with optical feedback: S.Sivaprakasam, E.M.Shahverdiev, P.S.Spencer and K.A.Shore, Phys.Rev.Lett **87**, 154101 (2001). We establish that our approach can explain the experimental results for *only* special cases.
PACS number(s):05.45.Xt, 05.45.Vx, 42.55.Px, 42.65.Sf

Seminal papers on chaos synchronization [1] have stimulated a wide range of research activity: a recent comprehensive review of the subject is found in [2].Synchronization phenomena in coupled systems have been especially extensively studied in the context of laser dynamics,electronic circuits, chemical and biological systems [3]. Application of chaos synchronization can be found in secure communication, optimization of nonlinear system performance, modeling brain activity and pattern recognition phenomena [2].

There are different types of synchronization in interacting chaotic systems. Complete, generalized, phase, lag and anticipating synchronizations of chaotic oscillators have been described theoretically and observed experimentally. Complete synchronization implies coincidence of states of interacting systems, $y(t) = x(t)$ [1]. A generalized synchronization, introduced for drive-response systems, is defined as the presence of some functional relation between the states of response and drive, i.e. $y(t) = F(x(t))$, [4]. Phase synchronization means entrainment of phases of chaotic oscillators, $n\Phi_x - m\Phi_y = const$, ($n$ and $m$ are integers) whereas their amplitudes remain chaotic and uncorrelated [5]. Lag synchronization appears as a coincidence of shifted-in-time states of two systems, $y(t) = x_\tau(t) \equiv x(t - \tau)$ with positive $\tau$ and has been studied in between symmetrically coupled non-identical oscillators [6] and in time-delayed systems [7,8]. Lastly, it was recently discovered [9] that dissipative chaotic systems with a time-delayed feedback can drive identical systems in such a way that the driven systems anticipate the drivers by synchronizing with their future states. Thus, anticipating synchronization [9] also appears as a coincidence of shifted-in-time states of two coupled systems, but in this case, in contrast to lag synchronization, the driven system anticipates the driver, $y(t) = x(t + \tau)$ or $x = y_\tau, \tau > 0$.

The first experimental observation of anticipating chaos synchronization has been reported recently in an optical system using bi-directionally coupled two diode lasers as transmitter and receiver [10]. The transmitter was rendered chaotic by application of an optical feedback in an external cavity configuration. The receiver was a solitary laser diode, i.e.with no feedback mirror. It was found that the anticipating time was equal to the light propagation time from the transmitter to the receiver and does not depend on the external cavity round trip time.

---

[1]Permanent address: Institute of Physics, 370143 Baku,Azerbaijan

In this paper we consider an analytical investigation of *exact* anticipating synchronization between two bi-idirectionally coupled non-chaotic time-delayed systems in the case when these systems are governed by two characteristic times: the delay time in the coupling and the delay time in the coupled systems. No approximations are to be made to achieve anticipating synchronization. For this reason we call it it exact anticipating synchronization. We find that exact anticipating chaos synchronization manifold with coupling-delay anticipating times can exist if there are parameter mismatches. We apply our approach to interpret the results of recent first experimental observation of anticipating chaos synchronization in external cavity semiconductor lasers with optical feedback. We establish that our approach can explain the experimental results *only* for special cases.

We begin with the investigation of non-chaotic systems. Although this is not the focus of the present work it is presented in detail because the non-chaotic systems serves as a test case where the idea of parameter mismatch leading to anticipating synchronization with coupling delay is tested to see which parameters' changes can be optimal, as, it will become clear below there are obvious qualitative similarities between the error dynamics equations for the non-chaotic and chaotic systems studied in this paper. Consider the following bi-directionally coupled non-chaotic time-delayed systems driver (x) and response (y) systems with the feedback delay time $\tau_1$ and the coupling delay time $\tau_2$:

$$\frac{dx}{dt} = -\alpha_1 x + k_1 x_{\tau_1} + k_3 y_{\tau_2},$$
$$\frac{dy}{dt} = -\alpha_2 y + k_2 y_{\tau_1} + k_3 x_{\tau_2}, \tag{1}$$

where $k_1$ and $k_2$ are feedback rates for the driver and response systems, respectively; $k_3$ is the coupling rate. Suppose that in eqs.(1) the parameter $\alpha$ is different for nonchaotic driver (master) and response (slave) systems: $\alpha_1 = \alpha + \delta$ and $\alpha_2 = \alpha - \delta$, where $\delta$ determines parameter mismatches. Now we argue that with parameter mismatches $\delta \neq 0$,

$$x = y_{\tau_2}, \tag{2}$$

is the anticipating synchronization manifold. The proof comes from the investigation of the dynamics of the error $\Delta = x - y_{\tau_2}$, which can be written as

$$\frac{d\Delta}{dt} = -\alpha_1 \Delta + k_1 x_{\tau_1} + k_3 y_{\tau_2} - k_2 y_{\tau_1+\tau_2} - k_3 x_{2\tau_2} - 2\delta y_{\tau_2}, \tag{3}$$

It is obvious that under the conditions

$$k_3 = 2\delta, \tau_1 = 2\tau_2, k_1 = k_2 + k_3, \tag{4}$$

equation (3) can be rewritten as

$$\frac{d\Delta}{dt} = -\alpha_1 \Delta + k_2 \Delta_{2\tau_2}. \tag{5}$$

In other words, under the conditions (4) $\Delta = 0$ is solution of the system (5).
To study the stability of the anticipating synchronization manifold (2) one can use a Krasovskii-Lyapunov functional approach. According to [11-12], the sufficient stability condition for the trivial solution $\Delta = 0$ of time delay equation $\frac{d\Delta}{dt} = -r(t)\Delta + s(t)\Delta_\tau$ is: $r(t) > |s(t)|$. Thus we obtain that

$$\alpha_1 > |k_2| \tag{6}$$

is the sufficient stability condition for the anticipating synchronization manifold (2). The condition (4) is the existence (necessary) condition of anticpating synchronization between the bi-idirectionally coupled non-chatotic time-delayed systems (1). Notice that we make no approximations to obtain the anticipating synchronization manifold (2) and this is why we address it as an *exact* anticpating synchronization. The severe resrtictiveness of the existence conditions (4) should be underlined, too. For the no-feedback case in the receiver, i.e. $k_2 = 0$ as in the experimental work [10], we obtain the following existence conditions for the exact anticipating synchronization manifold: $k_1 = k_3, \tau_1 = 2\tau_2$. Here we also have to underline the following point. Coupling-delay *exact* anticipating synchronization also can be observed for the periodic driving. Indeed, for the periodic driving

$$x(t) = x(t - 2\tau_2), \tag{7}$$

for

$$k_1 = k_2, \tag{8}$$

eq.(3) can be rewritten as:

$$\frac{d\Delta}{dt} = -(\alpha + k_3)\Delta + k_2 \Delta_{\tau_1}, \tag{9}$$

, which means that for the periodic driving the condition $\tau_1 = 2\tau_2$ is not required for the observation of the *exact* anticipating synchronization manifold (2). In general, *an approximate* antcipating synchronization manifold $x \approx y_{\tau_2}$ still holds, if the difference $k_1 x_{\tau_1} - k_3 x_{2\tau_2}$ is close to zero.

Next we consider anticipating synchronization between chaotic semiconductor lasers with optical feedback. Studying synchronization in chaotic semiconductor lasers is of great practical importance, due to their applications in high-speed optical communications and potential in chaos-based secure communications [7-8]. External cavity laser diodes are commonly modeled with the Lang-Kobayashi equations, see, e.g. [13]. Based on [14] here we will use rate equations for the laser intensity (the photon density) $I$ and excess carrier density $n$ to describe semiconductor lasers with optical feedback. The use of the rate equations, which neglect the optical phase is justified in detail in [14]. (We have also investigated the anticipating chaos synchronization regime in external cavity lasers modelled by the Lang-Kobayashi equations, which includes the optical phase and have derived the same existence conditions for the anticipating synchronization manifold as in the case of the simpler rate equations presented here.). Suppose that the master laser

$$\frac{dI_1}{dt} = (gn_1 - \gamma_1)I_1 + k_1 I_1(t - \tau_1) + k_3 I_2(t - \tau_2),$$

$$\frac{dn_1}{dt} = a - \gamma_e n_1 - gn_1 I_1, \tag{10}$$

is coupled bi-directionally with the slave laser

$$\frac{dI_2}{dt} = (gn_2 - \gamma_2)I_2 + k_2 I_2(t - \tau_1) + k_3 I_1(t - \tau_2),$$

$$\frac{dn_2}{dt} = a - \gamma_e n_2 - gn_2 I_2, \tag{11}$$

where $g$ is the differential optical gain; $\tau_1$ is the master and slave lasers' external cavity round-trip time; $\tau_2$ is the coupling delay time between lasers; $\gamma_e$- the carrier density rate; $\gamma_{1,2}$-the cavity decay rates(cavity losses): $\gamma_1 = \gamma + \delta$ and $\gamma_2 = \gamma - \delta$, where $\delta$ determines parameter mismatches; $a$-the injection (pump) currents for the lasers; $k_1$, $k_2$ and $k_3$ are feedback and coupling rates. For simplicity we take the injection current the same for both lasers. In the following we will consider *only* the case of solitary receiver, as in the experimental work [10]. Now we analytically prove that under certain conditions the manifold

$$I_1 = I_{2,\tau_2}, n_1 = n_{2,\tau_2}, \tag{12}$$

can be the *exact* anticipating synchronization manifold for the interacting laser systems (10-11) and derive those conditions. Let $\Delta_1$, $\Delta_2$ be the deviations from the synchronization manifold (12): $\Delta_1 = I_1 - I_{2,\tau_2}$ and $\Delta_2 = n_1 - n_{2,\tau_2}$. Then quite easily one can obtain the following equation for the joint error dynamics for $\Delta_1$ and $\Delta_2$:

$$\frac{d\Delta_1}{dt} + (\gamma + \delta)\Delta_1 - k_1 I_{1,\tau_1} - k_3 I_{2,\tau_2} + k_3 I_{1,2\tau_2} + 2\delta I_{2,\tau_2} = -\frac{d\Delta_2}{dt} - \gamma_e \Delta_2, \tag{13}$$

. Thus under the conditions

$$k_1 = k_3 = 2\delta, \tau_1 = 2\tau_2 \tag{14}$$

$\Delta_1 = \Delta_2 = 0$ is the solution of equation (13), in other words we obtain the *exact* anticipating synchronization manifold (12). We underline that while deriving (13) we did not assume that the $\Delta_1$ and $\Delta_2$ are small. Also notice that without parameter mismatches one cann't obtain the *exact* anticipating chaos synchronization manifold (12). The condition (14) is the necessary condition for *exact* anticipating synchronization between bi-directionally coupled laser systems(10-11).

There is an important point which deserves to be underlined. Had we retained the feedback term in the receiver system, i.e. $k_2 \neq 0$, then all the discussions relating to the formulai (3-9) for the existence conditions for coupling- delay anticipating synchronization for the non-chaotic systems would be fully applicable for systems (10) and (11). Again as in the case of non-chaotic systems, in general, the *approximate* antcipating synchronization manifold $I_1 \approx I_{2,\tau_2}, n_1 \approx n_{2,\tau_2}$ can still take place under less srticter conditions than (14). The differences between the pump currents for the lasers can even be helpful to achieve *an approximate* anticipating synchronization.

Thus the theory of *exact* anticipating synchronization for bi-directionally coupled external cavity laser diodes with parameter mismatches (we also considered possible mismatches of other parameters $g$ and $\gamma_e$, but we obtained better results with mismatches of $\gamma$) considered here only *partially* explains experimental results on anticipating chaos synchronization in [10]. The specific value of external cavity round trip time $\tau_1 = 2\tau_2$ for which we obtain *exact* anticipating chaos synchronization for the bi-directionally coupled laser systems does not mean that one can rely on the numerical results in [13] where anticipating chaos synchronization between the unidirectionally coupled laser diodes was investigated and the anticipating synchronization manifold $I_1 = I_{2,\tau_1-\tau_2}$ was established. This conclusion follows from the fact that in general $I_1 = I_{2,\tau_1-\tau_2}$ is not the anticipating synchronization manifold for systems (10) and (11).

To summarize, we have presented an analytical investigation of exact anticipating synchronization between two bi-idirectionally coupled time-delayed systems in the case when these systems are governed by two characteristic times: the delay time in the coupling and the delay time in the coupled systems themselves. We have found that exact anticipating chaos synchronization with coupling-delay anticipating times can exist if there are parameter mismatches. We have applied our approach to interpret the results of recent first experimental observation of anticipating chaos synchronization in external cavity semiconductor lasers with optical feedback. We have shown that our approach can explain the experimental results *only* for special cases.

This work is supported by UK EPSRC under grants GR/R22568/01 and GR/N63093/01.